# Design, Construction and Commissioning of the Digital Hadron Calorimeter - DHCAL

C. Adams[a], A. Bambaugh[d], B. Bilki[a,e], J. Butler[c], F. Corriveau[f], T. Cundiff[a], G. Drake[a], K. Francis[a,*], B. Furst[a], V. Guarino[a], B. Haberichter[a], E. Hazen[c], J. Hoff[d], S. Holm[d], A. Kreps[a], P. DeLurgio[a], Z. Matijas[a], L. Dal Monte[d], N. Mucia[g], E. Norbeck[e,†], D. Northacker[e], Y. Onel[e], B. Pollack[g], J. Repond[a], J. Schlereth[a], F. Skrzecz[a], J. R. Smith[a,h], D. Trojand[f,#], D. Underwood[a], M. Velasco[g], J. Walendziak[a], K. Wood[a], S. Wu[c], L. Xia[a], Q. Zhang[a,b,&], A. Zhao[a]

[a]*Argonne National Laboratory, 9700 S. Cass Avenue, Argonne, IL 60439, U.S.A.*
[b]*Institute of High Energy Physics, Chinese Academy of Sciences, P.O. Box 918, Beijing, 100049, PRC*
[c]*Boston University, 590 Commonwealth Avenue, Boston, MA 02215, U.S.A.*
[d]*Fermilab, P.O. Box 500, Batavia, IL 60510-0500, U.S.A.*
[e]*University of Iowa, Iowa City, IA 52242-1479, U.S.A.*
[f]*McGill University, 3600 University Street, Montreal, QC H3A 2T8, Canada*
[g]*Northwestern University, 2145 Sheridan Road, Evanston, IL 60208-3112, U.S.A.*
[h]*University of Texas at Arlington, P.O. Box 19059, Arlington, TX 76019, U.S.A.*
* Now at Northern Illinois University
# Now at Windsor University
& Now at Xi'an Jiaotong University
†Deceased

**Abstract.** A novel hadron calorimeter is being developed for future lepton colliding beam detectors. The calorimeter is optimized for the application of Particle Flow Algorithms (PFAs) to the measurement of hadronic jets and features a very finely segmented readout with $1 \times 1$ cm$^2$ cells. The active media of the calorimeter are Resistive Plate Chambers (RPCs) with a digital, i.e. one-bit, readout. To first order the energy of incident particles in this calorimeter is reconstructed as being proportional to the number of pads with a signal over a given threshold. A large-scale prototype calorimeter with approximately 500,000 readout channels has been built and underwent extensive testing in the Fermilab and CERN test beams. This paper reports on the design, construction, and commissioning of this prototype calorimeter.

# INTRODUCTION

At a future lepton collider, such as the ILC or CLIC [1], the measurement of hadronic jets will play an important role in discovering or exploring physics beyond the current Standard Model of Particle Physics. Indeed, both the energy resolution and the mass resolution of multi-jet systems will be important in defining the physics reach of this new facility. Of particular interest will be the identification of electroweak bosons through their hadronic decay. Identification of $Z^0$ and $W^\pm$ on an event-by-event basis will require an energy resolution of the order of 3-4% for a wide range of jet energies.

A novel approach, named Particle Flow Algorithms (PFAs) [2] is proposed to achieve this unprecedented jet energy resolution. PFAs attempt to measure each particle in a hadronic jet individually, using the component providing the best energy/momentum resolution. In this approach, charged particles are measured with a high-precision tracker, photons with the electromagnetic calorimeter and the remaining neutral hadronic particles in a jet with the combined electromagnetic and hadronic calorimeters. Table I shows the average fraction of the jet energy carried by these particle types and the expected single particle resolution obtained with the appropriate detector subsystem.

**Table I.** Particles in a jet, their average fraction of the energy of a given jet and their contribution to the overall resolution.

| Particle | Average fraction of jet energy [%] | Measured with | Contribution to resolution [$\sigma^2$] |
|---|---|---|---|
| Charged | 65 | Tracker | Negligible |
| Photons | 25 | ECAL with 15%/$\sqrt{E}$ | $0.07^2 E_{jet}$ |
| Neutral hadrons | 10 | Calorimeter with 50%/$\sqrt{E}$ | $0.16^2 E_{jet}$ |
| **Total** | **100** | | **$0.18^2 E_{jet}$** |

Assuming typical electromagnetic and hadronic calorimeters with resolutions of 15%/$\sqrt{E}$ and 50%/$\sqrt{E}$, respectively, the overall jet energy resolution is predicted to be a spectacular 18%/$\sqrt{E}$ and to be dominated by the calorimetric measurement of the neutral hadrons. However, this estimation assumes that the energy deposits in the calorimeter can be associated unambiguously to charged particles (and therefore be ignored) or to neutral particles (which are measured with the calorimeter). Naturally, this will not always be possible and, indeed, minimizing the 'confusion term' due to misassignments of energy deposits constitutes the major challenge of this approach and imposes certain requirements on the detector design and in particular on the design of the calorimeters. A detector optimized for the application of PFAs requires an excellent tracker within a high magnetic field, a large inner radius of the calorimeter (to increase the distance between showers from different particles), the calorimeters to be placed inside the coil (to avoid energy losses in the inert material of the solenoid), calorimeters with extremely fine segmentation of the readout (to measure showers individually), an electromagnetic calorimeter with a short Molière radius (to reduce the lateral spread of electromagnetic showers) and a hadronic calorimeter with a short interaction length (to be to be placed inside a coil with the smallest possible inner radius).

In this context the CALICE collaboration [3] developed a Digital Hadron Calorimeter (DHCAL) based on Resistive Plate Chambers (RPCs). A large scale prototype, the DHCAL, was built in 2008 – 2010 and was subsequently tested in the Fermilab and CERN test beams [4-6]. This paper describes the design, construction and commissioning of the DHCAL prototype. An overview of the various tests beam activities is also given. A follow-up paper describes the details of the electronic readout system [7].

## DESCRIPTION OF THE DHCAL

The DHCAL prototype constitutes the first large-scale hadron calorimeter with digital readout and embedded front-end electronics. It also utilized, for the first time, a pad-readout together with RPCs. The design of the DHCAL was based on preliminary work done with a small-scale prototype, see references [8 – 13] for details. The DHCAL was exposed to particle beams both at Fermilab and at CERN.

At Fermilab, the DHCAL prototype consisted of two parts: 1) a 38-layer structure with 17.4 mm thick steel absorber plates and 2) a 14-layer structure with eight 2.0 cm thick steel plates followed by six 10.0 cm thick steel plates. The former (latter) is commonly referred to as the Main Stack (Tail Catcher and Muon Tracker (TCMT)). Both absorber structures were equipped with RPCs as active elements. Each layer measured approximately $1 \times 1$ m$^2$ and was inserted between neighboring steel absorber plates. The Main Stack rested on a movable stage, which offered horizontal and vertical movements in addition to the possibility of rotating the entire stack on a vertical axis [14]. The distance from layer-to-layer in the 38-layer structure was 3.17 cm. To allow for rotation of the Main Stack, the first layer in the TCMT was spaced 40.1 cm downstream of the last layer in the Main Stack.

In addition to this standard configuration, data were also collected with 50 layers in a structure without absorber plates. In this configuration the skins of the detector cassettes (2 mm copper and 2 mm steel) and the glass of the RPCs served as the only absorber material. Here the layers were spaced 2.54 cm apart. Figure 1 shows photographs of the Main Structure, the TCMT, and of the configuration with minimal absorber material.

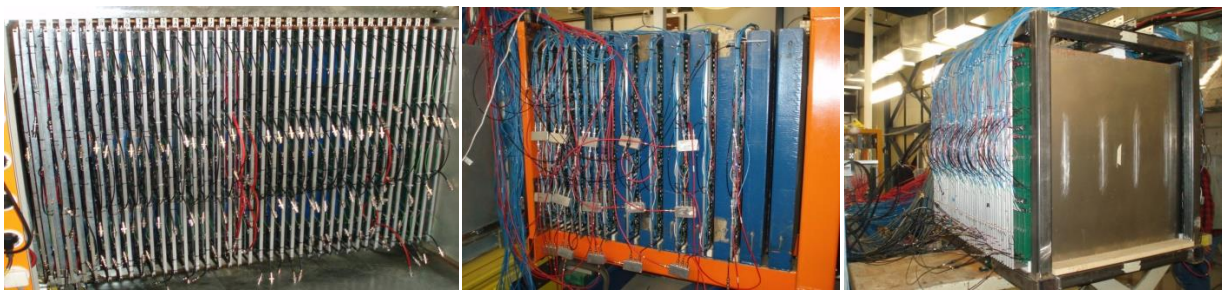

**Figure 1.** Left: The Main Stack (before cabling), middle: the TCMT, right: the 50-layer configuration with minimal absorber material.

In 2012 the DHCAL layers were shipped to CERN to be tested with an absorber structure containing 39 tungsten plates, followed by a steel-plate TCMT identical in design to the one used at Fermilab. The tungsten plates were made of an alloy with 93% tungsten and a balance of copper and nickel. The plates measured 10 mm in thickness, which corresponds to about 2.6 radiation lengths $X_0$ and 0.09 nuclear interaction lengths $\lambda_I$. For easier transportation between test beams, the Main Stack, the back-end readout electronics, and the high-voltage power supply system were all placed on a platform. A photograph of the setup is shown in Fig. 2.

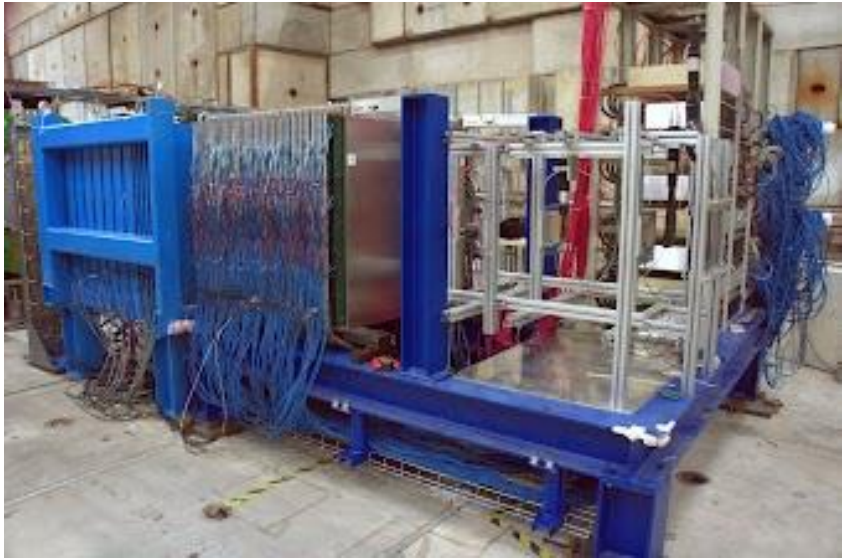

**Figure 2.** Photograph of the DHCAL setup at CERN showing the Main Stack with tungsten plates followed by the TCMT with steel plates.

Each layer consisted of three RPCs, each with an area of $32 \times 96$ cm$^2$. The three RPCs were stacked vertically to create an active area of approximately $1 \times 1$ m$^2$. Each RPC in turn was read out with two Readout boards, which covered the entire gas volume of the chambers. The three chambers and their boards were contained in a cassette structure providing the necessary mechanical protection during transportation and installation.

Each RPC was read out by 3072 $1 \times 1$ cm$^2$ pads located on the bottom side of the Readout boards. To first order the charge generated by a single avalanche in an RPC is solely determined by the ionization in the gas gap that is the furthest away from the anode and thus subject to the largest gas multiplication. This leads to a wide range of signal charges [8] which are, if at all, only weakly correlated to the energy loss in the gas gap. Therefore, a precise measurement of the induced charge is seen as of only limited interest. As a consequence, the DHCAL readout was simplified to a single threshold per pad. This approach is commonly denominated as a digital readout. In the configuration with $39 + 15 = 54$ active layers, the channel count was $54 \times 96 \times 96 = 497,664$, which at the time constituted a world record in calorimetry.

Table II summarizes the various configurations tested in test beams and lists their relative thicknesses in radiation ($X_0$) and nuclear interaction lengths ($\lambda_I$). Notice the large difference in thickness in radiation lengths of the Main Stack at Fermilab and CERN, despite their similar thicknesses in interaction lengths.

**Table II.** Summary of the various DHCAL configurations and their thicknesses in radiation lengths $X_0$ and nuclear interaction lengths $\lambda_I$.

| Location | Structure | Number of layers | Thickness in $X_0$ | Thickness in $\lambda_I$ |
|---|---|---|---|---|
| FNAL | Main stack (Fe) | 38 | 48.6 | 5.23 |
| | TCMT | 14 | 47.3 | 5.00 |
| | **Total** | **52** | **95.9** | **10.23** |
| CERN | Main stack (W) | 39 | 112.7 | 4.84 |
| | TCMT | 15 | 53.3 | 5.63 |
| | **Total** | **54** | **166.0** | **10.47** |
| FNAL | Minimal absorber | 50 | 14.4 | 1.69 |

# DESIGN, CONSTRUCTION AND OPERATION OF RPCs

Each detector layer consisted of three RPCs with the approximate dimensions of $32 \times 96$ cm$^2$. The resistive plates were made of soda-lime glass with a bulk resistivity of about $4.7 \cdot 10^{13}$ $\Omega$cm. To minimize the distance between the gas gap and the anode readout and thus to minimize the average pad multiplicity for minimum ionizing particles traversing the chamber [8], the thickness of the anode plate was chosen to be 0.85 mm, which was considered to be close to the minimum value acceptable for safe mechanical handling. The cathode plates were chosen to be slightly thicker, i.e. 1.15 mm. One side of each glass plate was coated with resistive paint, as will be described below.

The sides of the chambers were assembled using custom-made extruded channels made of PVC. A cross section of such a channel is shown in Fig. 3. With the help of these channels and nylon strings (fishing lines) partially clad in PVC sleeves the thickness of the gas gap was maintained uniformly over the surface of the chambers. The PVC sleeves increased the diameter of the fishing lines to match the corresponding dimension of the PVC channels and also served to create a chicane for the gas flow, as seen in Fig. 4. The chicane ensured that the gas flow was uniform in all regions of the chamber, thus minimizing the required gas flow to maintain an optimal performance of the chambers.

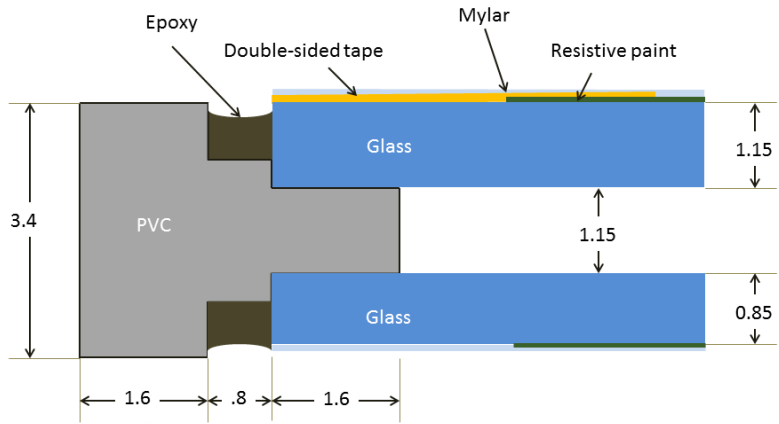

**Figure 3.** Cross section through the PVC channels used as outer rims for the RPCs. The drawing is not to scale, and the units are in [mm].

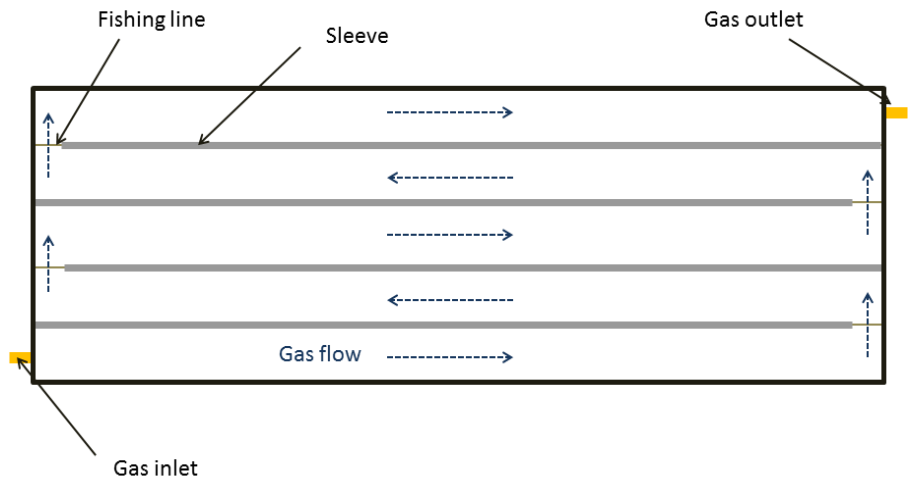

**Figure 4**. Top view of an RPC showing the layout of the fishing lines and their sleeves. The gas flow is indicated with dashed arrows. The drawing is not to scale.

To apply high voltage across the gas volume, one side of each glass plate was coated with resistive paint. The ideal surface resistivity of the coat was determined to be $1 - 5$ M$\Omega/\square$, where larger values decrease the rate capability of the chambers and lower values, in particular on the anode side, increase the average pad multiplicity [8]. After masking off a rim of $3 - 4$ mm from the edges, the glass plates were coated with a two-component 'artist' paint[1]. This non-toxic paint was applied with a spray gun mounted on a computer-controlled arm, as shown in Fig. 5. The whole operation was located in a special booth with an exhaust fan to remove the fumes generated during the spraying operation. Apart from a significant set-up and cleaning time, the application took approximately two minutes per plane.

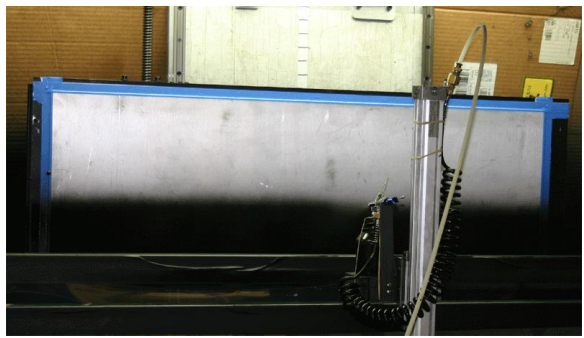

**Figure 5.** Photograph showing the spraying of a sheet of glass with resistive paint.

After the paint had dried, the surface resistivity was measured on a grid over the entire surface of the glass. Figure 6 shows an example of the obtained uniformity. While the uniformity over the surface was in general adequate, large variations of the average value from plate to plate were observed. As a result, only 60% of the sprayed plates could be used for constructing RPCs, the others having to be cleaned and re-sprayed. The reason for the observed large variation in resistivity was most likely due to the typically large fluctuations in environmental conditions in the Chicago area and the rudimentary control over these conditions exerted during the spraying operation. In total over 700 glass plates were sprayed. Of the ones used, the average surface resistivity measured 3.8 MΩ/□ with a standard deviation of approximately 20%.

---

[1] The two components were Chrome Oxide Green 8-40 and Air-Opaque Black paint from Badger Air-brush Company, Franklin Park, Il 60131. The two components were mixed in roughly the same amounts. The amounts were fine tuned depending on the outcome of the spraying of the previous set of glass plates.

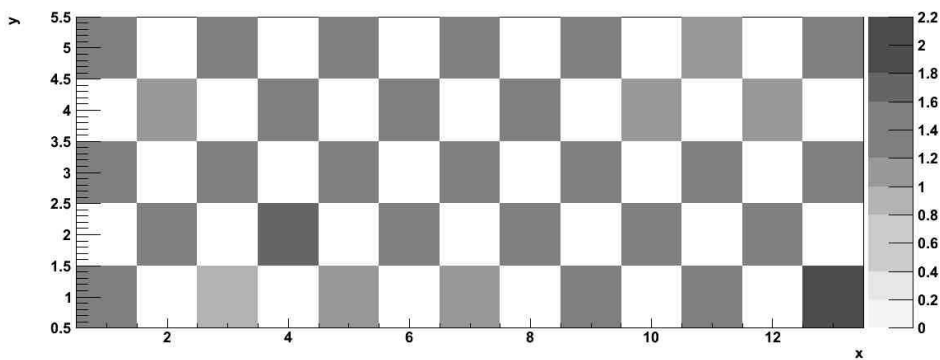

**Figure 6.** Measured resistivity of a painted glass sample. Values are in MΩ/□. The areas in white were not measured.

The assembly of the chambers involved the following sequence of steps: the PVC frame pieces were cut to size and were glued together at their edges using a custom gluing fixture. This provided the outer rim of the chamber. To ensure the best possible uniformity of the gas gap, the gluing fixtures were machined to a precision of 25 – 50 μm. Next a sheet of the thicker glass was placed on the frame with the painted side facing upwards, and the gaps between the glass plates and the rim (see Fig.3) were filled with Epoxy to achieve the required rigidity and create a seal to contain the gas. After allowing the glue to dry overnight, the frame and glass were flipped over. Four nylon fishing lines, covered to approximately 95% of their length with plastic sleeves (1.15 mm diameter) were inserted across the length of the glass and were slightly tensioned. A sheet of the thinner glass, again with the painted side facing upwards, was placed over the frame and glued in place. Once the glue was cured, the gas inlet tubes, high voltage mounting fixtures and the fishing lines were glued in place. The chambers on both the high-voltage and the ground side were protected with a sheet of Mylar each with a thickness of 50 μm.

In total 205 RPCs were produced at three different assembly stations. The assembly of each chamber required about 1.5 man-days of effort. Figure 7 (left) shows a photograph of one of the three assembly stations, including a partially completed chamber with the fishing lines/gas spacers installed. A completed chamber is shown in Fig.7 (right).

To be used in calorimetry, the chambers needed to be efficient over the entire surface, including the outer edges of the chambers. To this effect, the resistive paint was applied as close as possible to the edge of the glass plates. As a result, some chambers exhibited a significant amount of current around the edges, in worst cases seen as a glowing in the dark. In order to increase the resistivity and reduce such currents, a narrow strip of double-sided tape was applied on the high voltage side of the chambers along the entire edge before covering the chamber with Mylar, see Fig. 3. This reduced the dark current significantly. With the additional tape, the chambers typically drew a dark current of 0.1-0.2 µA when operated at 6.3 kV.

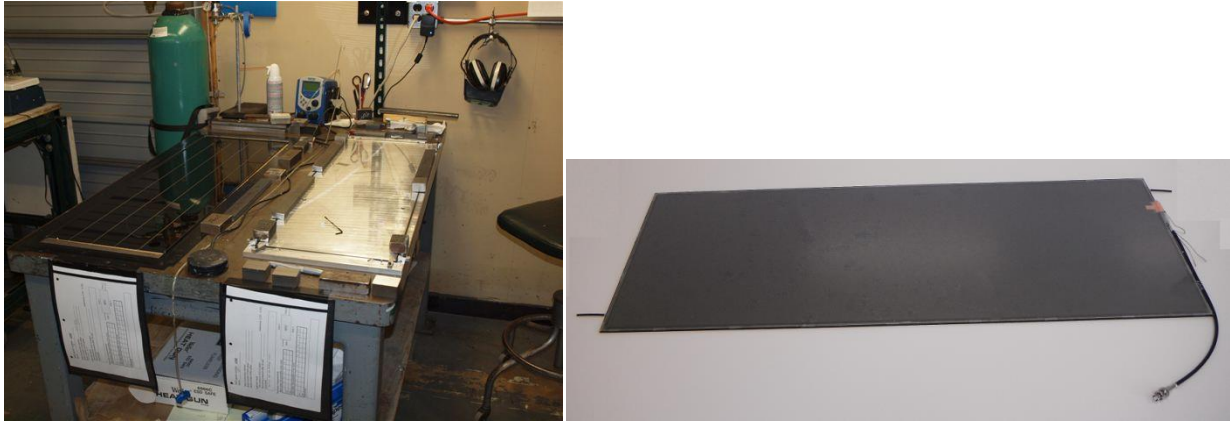

**Figure 7.** Left: one of the three assembly stations for the construction of RPCs. Right: a completed chamber.

The RPCs were operated in avalanche mode with a default high voltage of 6.3 kV. The gas was a mixture of three components Tetrafluorethane (R134A) : Isobutane : Sulfurhexafluoride ($SF_6$) with the following ratios 94.5 : 5.0 : 0.5. This mixture is non-flammable. At CERN, due to the higher location, the default high voltage was reduced to 6.1 kV, providing a similar performance (efficiency and average pad multiplicity) as at Fermilab.

## QUALITY ASSURANCE DURING THE ASSEMBLY OF RPCs

To test for gas tightness, completed chambers were slightly compressed (using weights) to avoid bulging and their gas volume was put under pressure. The pressure was slowly increased to a maximum of 75 Pa. The chamber passed the test if the pressure drop was less than 5 Pa in 30 seconds. Chambers not passing the first test were repaired and subsequently retested. All chambers eventually passed the pressure test.

To assure uniformity in the gap, the thickness of a sample of chambers was measured along their edges. Since the glass plates themselves were highly uniform in thickness, this measurement provided a direct measure of the gap size along the edges. In general these were found to be within ±0.1 mm, see Fig. 8. Corners were often somewhat thicker (up to 0.3 – 0.4 mm), but this affected the performance of the chambers only in a very small area. Five RPCs were found to have low efficiency regions at the corners or along the sides, due to unusually large gaps at those places. These chambers were eventually removed from the stack and replaced with chambers with more uniform gap sizes and efficiencies/pad multiplicities. Due to the fishing lines, the central regions of the chambers are assumed to be uniform by construction and their thickness were not measured.

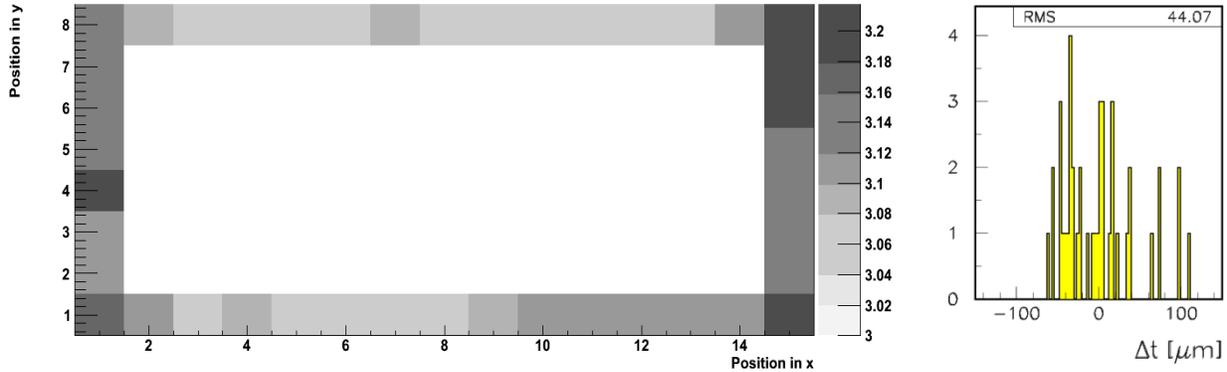

**Figure 8.** Left: Measurement of the thickness (units are in mm) along the edges (positions are in arbitrary units) of a typical chamber. Right: Distribution of the differences from the mean thickness for some of the chambers.

Every chamber was tested at 7.0 kV (the default operating voltage being 6.3/6.1 kV) for approximately 24 hours. Only chambers maintaining a current less than 0.5 µA were accepted for installation into the stack. Further tests were performed using cosmic rays as described below.

# ELECTRONIC READOUT SYSTEM

A detailed description of the electronic readout system will be published in a follow-up paper [14]. The following chapter summarizes the main features of the system.

### a) Overview of the electronic readout system

The electronic readout system was optimized for the readout of a large number of channels and offers the option to be scaled up to the tens of millions of readout channels envisaged for a hadron calorimeter operating at a future lepton collider. A block diagram of the system is shown in Fig. 9. The electronics is divided into two parts: The "on-detector" electronics processes charge signals from the detector, collects data for transmission out, and acts as the interface for slow controls. The "back-end" electronics receives and processes the streams of data from the front-end electronics, and in turn passes it to the Data Acquisition (DAQ) system. It also has an interface to the timing and trigger systems.

Because of the high channel count, a custom integrated circuit has been developed for the front-end electronics. The device, called DCAL, performs in addition to ancillary control functions, all of the front-end processing, including signal amplification, discrimination/comparison against threshold, recording the time of the hit, temporary storage of data, and data read out. It services 64 detector channels with a choice of two programmable gain ranges with sensitivities of ~10 fC and ~100 fC, respectively.

When a charge is received by a front-end amplifier and it exceeds the programmable threshold (common to all channels in a chip), the hit pattern of all channels in the chip is recorded along with the time. The timing of hits in the DCAL chip is implemented using the concept of a "timestamp" counter. This counter is reset once per second across the system and advances with each 100 ns clock. The latter is synchronous across the entire system. The data are captured in a readout buffer inside the chip either from an external trigger or self-triggered and are read from the chip using high-speed serial bit transmission. The chip also has slow control functions, on-board charge injection, and the ability to mask off noisy channels.

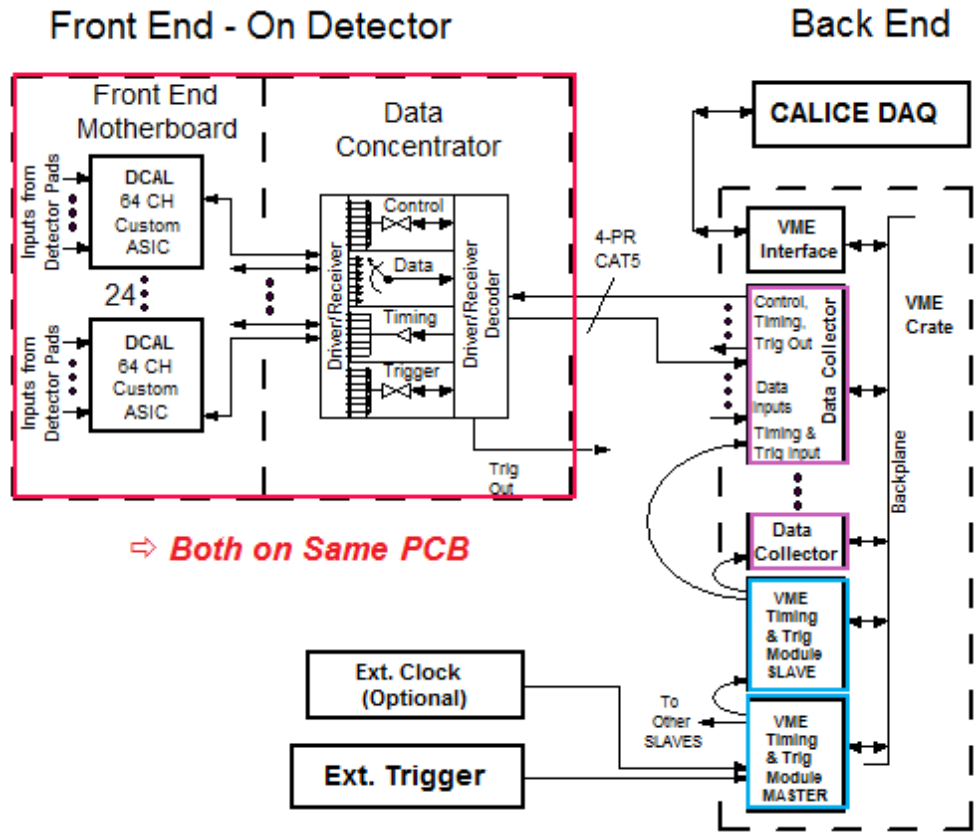

**Figure 9.** Block diagram of the electronic readout system.

The chips reside on sophisticated Front-end printed circuit boards that are part of the Readout board. Each Readout board consists of a Pad board, containing the $1 \times 1$ cm$^2$ pads, and the Front-end board. Charge signals pass from the pads on the bottom side of the Pad board to the input pins of the DCAL chips that reside on the top side of the Front-end board. The inner layers of the Front-end board also contain the routing of the clock and digital control and data lines, as well as power and ground. Each Front-end board contains 24 chips, servicing 1,536 individual channels. There is virtually no dead detector space due to the electronics, and the Readout boards can be tiled on three sides to cover large detector planes.

The output data streams from the DCAL chips send their data to the Data Concentrator Field Programmable Gate Array (FPGA) located on the narrow edge of the Front-end board. The data coming out of the Data Concentrator is both time-ordered and zero-suppressed. The Data Concentrator also serves as interface for sending slow control information to the DCAL chips using a serial bus protocol.

The readout operation overlaps with signal acquisition, making noise performance critical. The data is read out serially from the Front-end boards using "data push" into custom VME cards in the back-end system called Data Collectors. The data are time-sorted using the timestamps and are stored in readout buffers. The data are read periodically into a computer, where higher-level algorithms perform the event reconstruction. In addition, the Data Collectors provide an interface to the front-ends for slow control communication and timing.

The VME crates that host the Data Collector also contain a Timing & Trigger Module that receives timing and trigger signals from peripheral subsystems and communicates with the Data Collectors to provide this information to the front-ends.

Additional details on the electronic readout system can be found in [14].

### b) Gluing of the Front-end boards and Pad-boards

The Read-out board is a sandwich, comprised of the Front-end board and the Pad board and bonded together using conductive epoxy. This solution was adopted to avoid the use of blind or buried vias, which would have been difficult to implement in a board of this size, i.e. $32 \times 48$ cm$^2$.

The Front-end board features eight layers using conventional fabrication methods and materials. The Pad board is a 2-layer board, with a single via connection between each of the $1 \times 1$ cm$^2$ pads on the bottom and the $0.5 \times 0.5$ cm$^2$ glue pads on the top. The via holes are filled with silver epoxy to provide a smooth surface on the pad side.

A robotic gluing machine dispensed glue dots onto the glue pads of the Pad board, as shown in Fig. 10. A two-component epoxy, Epo-tek E4110[2], was chosen for its relatively long pot life of four hours. After dispensing, the fully assembled and tested Front-end board was placed onto the Pad board and weighted down using metal blocks. The combined Readout boards were then heated in an oven to allow the glue to fully cure. After the initial learning curve, the success rate of the gluing process was in excess of 99%. Figure 11 shows measurements of the resistivity between a pad on one side of the Readout board and the corresponding input to the DCAL chip. The measured values include an offset of approximately 0.1 Ω due to the leads of the voltmeter. After performing a few such measurements and failing to identify a single faulty connection showing high resistivity, these measurements were discontinued for the remaining production of Readout boards.

# DESIGN AND CONSTRUCTION OF THE CASSETTES

The cassettes provided mechanical stability and protected the three RPCs and the six readout boards in a given layer from harm during both transportation to and insertion into the absorber structure. The front-plate consisted of 2 mm steel, while the back-plate was made of 2 mm copper. The latter was in thermal contact with the front-end chips of the Readout boards and helped dissipate the heat generated by the electronics. Figure 12 shows photographs of a partially and fully assembled cassette.

---

[2] Supplied by EPOXY TECHNOLOGY INC, Billerica, MA 01821

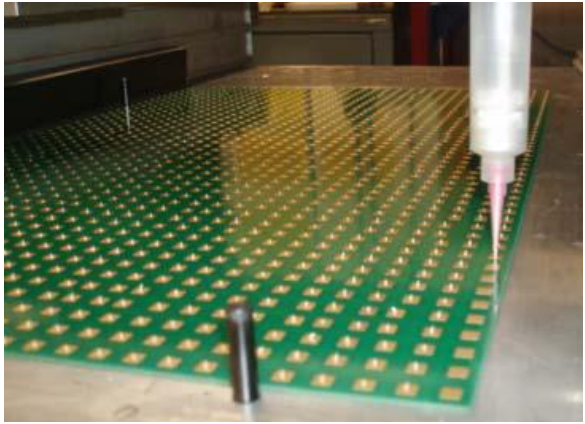

**Figure 10.** Glue machine dispensing glue onto the Pad-board.

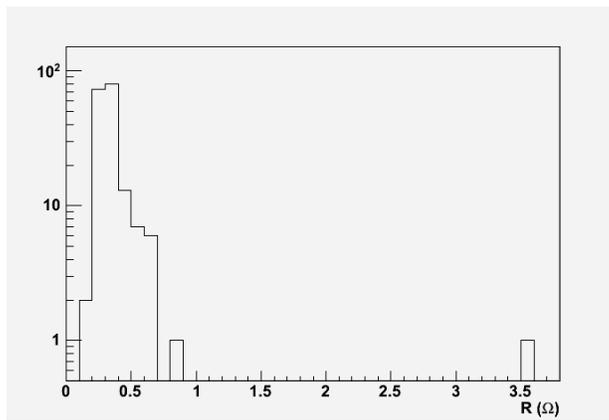

**Figure 11.** Measurements of the resistivity between pads on the Pad board and the corresponding inputs on the DCAL chip.

The average pad multiplicity depends critically on the distance between the gas gap and the readout pads. The cassettes were designed such that the Readout boards are flush with the Mylar protecting the resistive paint of the chambers, thus minimizing the distance of the readout pads to the gas gap. Both the front and the back plates of the cassettes contained a set of 4 × 2 matching holes positioned such as to correspond to the gap between RPCs, see Fig. 13. A badminton string was threaded through these holes, going from the front to the back between the lower two RPCs, and up to the gap between the two upper RPCs, where it was threaded back to the front. A knot at one end prevented the string to slip through the hole. The other end of the string was tensioned to about 10 kg and a second knot was applied. This resulted in a gentle pressure of the copper plate against the DCAL chips, which then transferred the pressure from the Readout board onto the Mylar of the RPC, thus minimizing potential gaps between the readout pads and the chambers.

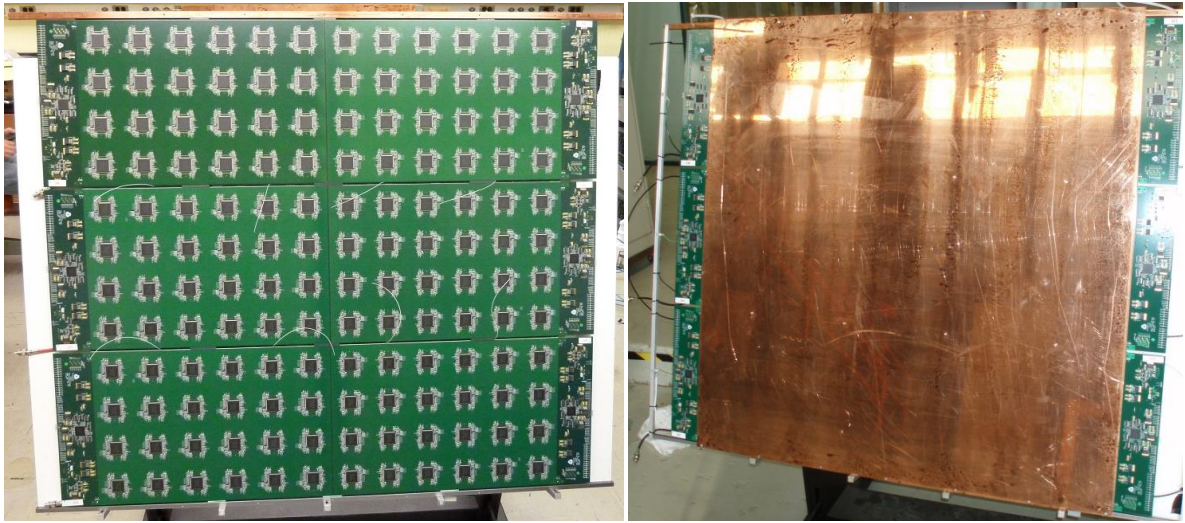

**Figure 12.** Photographs of a cassette without (left) and with (right) the copper back-plate.

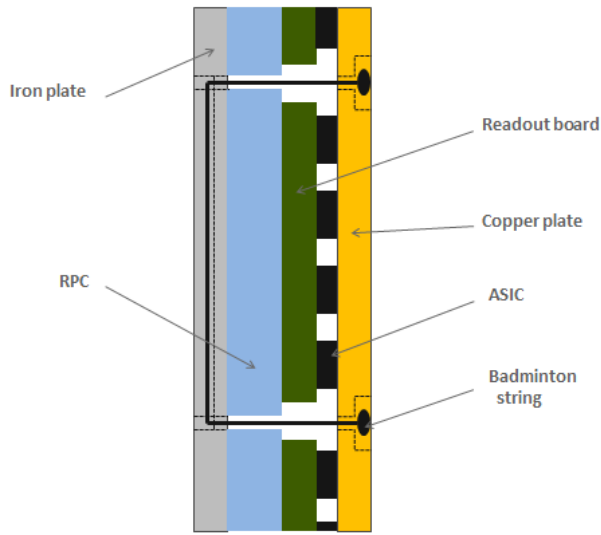

**Figure 13.** Cross-section through the cassette showing the compression device (not to scale).

Typically, each cassette took 45 – 60 minutes to assemble. The completed cassettes were flushed with gas and tested with the nominal high voltage. Before transportation to the Fermilab test beam, a noise run taken in trigger-less mode ensured the full functioning of the layer.

Particular care was devoted to develop a grounding scheme minimizing electronic noise which otherwise could produce accidental hits. Figure 14 shows a schematic of the grounding. Note that the ground at the low voltage power supplies was floating.

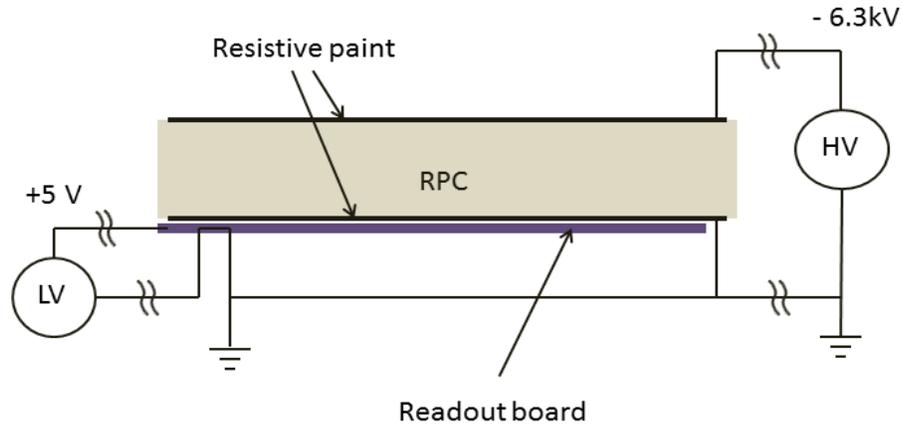

**Figure 14.** Grounding scheme of a DHCAL detector cassette.

# PERIPHERALS

Operating the DHCAL required three supplies: the appropriate gas mixture, high voltage for the RPCs and low voltage for the Front-end boards. In the following these supplies are briefly described.

The DHCAL gas system consisted of a mixing unit and a distribution rack. Both units were custom made. The mixing unit was supplied with the three gases to be mixed for operating the RPCs, i.e. Tetrafluorethane(R134A) : Isobutan : Sulfurhexafluoride ($SF_6$) = 94.5 : 5.0 : 0.5 %. The desired gas mixture was obtained with three electronic mass flow controllers made by Brooks Instruments [15]. The integral flow rate was 200 – 300 cm$^3$/minute. The distribution rack received the mixed gas and distributed it to up to 28 individual lines. Each line was equipped with a valve and a flow meter. To gauge the gas flow through a particular line, the return lines were equipped with bubblers. Typically, the input gas line was split into two to serve the six RPCs contained in two adjacent layers. The three RPCs in a layer were connected in series with the lower RPC receiving the input line and the uppermost RPC being connected to the exhaust line. The exhaust gas from all lines was combined into a single outlet and was vented into the atmosphere. Figure 15 shows photographs of the gas mixing and distribution racks.

The default high voltage for operating the RPCs was selected to be -6.3 kV, which provides high efficiency for detecting minimum ionizing particles, but is still well below the onset of streamers. Three High Voltage supply systems were employed during the various test beam campaigns (see below): a) a LeCroy 4032A system, b) Bertan units (NIM based), and c) a Wiener MPOD_2H mainframe together with 8 ISEG_ESH_80 80x_105 modules. The LeCroy systems were built in the late 70s/early 80s and proved to be unreliable, with frequent breakdowns. The Bertan units on the other hand, even though from the same time period, performed reliably. Their major drawback, however, was the lack of computer control. The later test beam campaigns used the modern Wiener-based system, which performed to satisfaction.

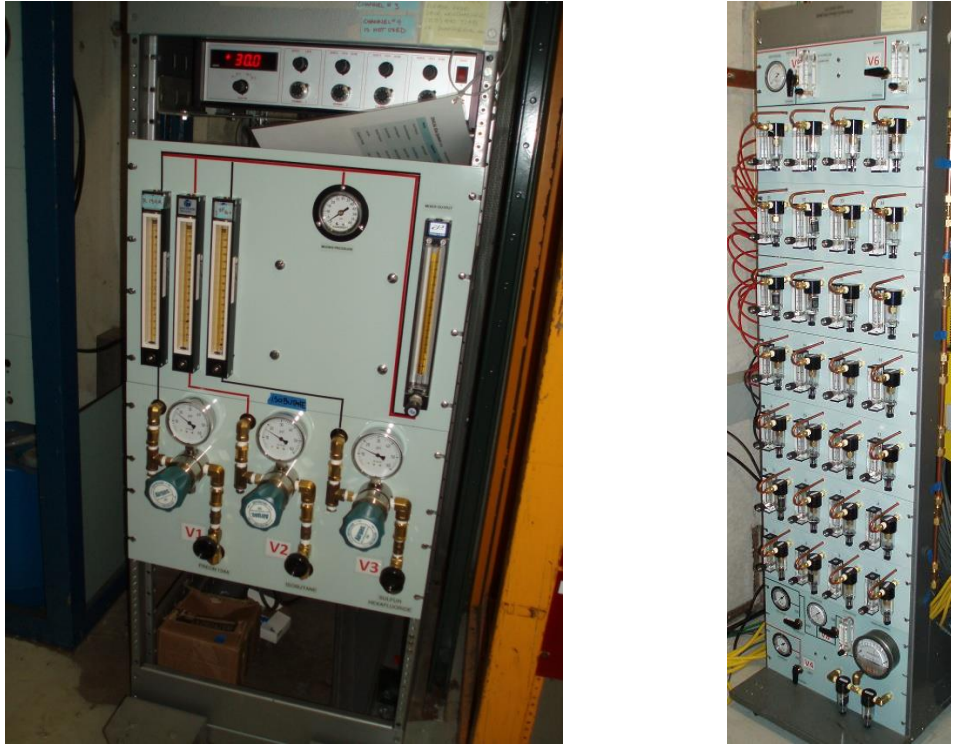

**Figure 15.** The DHCAL gas mixing system (left) and the gas distribution system (right).

In general, three RPCs in a layer were connected to one high voltage channel, the line being fanned out by a passive splitter. As the tests progressed, some chambers required higher voltages to operate with acceptable efficiency. Typically these chambers were then powered by their own individual line.

A comparison of the noise rate in the RPCs showed no difference between the three high voltage supply systems. Typical currents were 1 – 2 µA per layer, compared to the currents of 0.1 – 0.2 µA measured in the Cosmic Ray test stand. The increased currents were caused by the higher temperature of the chambers when operated inside an absorber structure.

The Readout boards required +5 V to operate. The voltage was provided by a set of seven Wiener PL508 power supplies. The power was distributed to the 300+ readout boards of the DHCAL using custom made distribution boxes, with individual switches and circuit breakers on each line. Each such box provided power to eight layers or 48 readout boards. The power per board was of the order of 15 Watts. Figure 16 shows a photograph of a rack containing five power supply/distribution box pairs. In total eight such pairs were acquired/built.

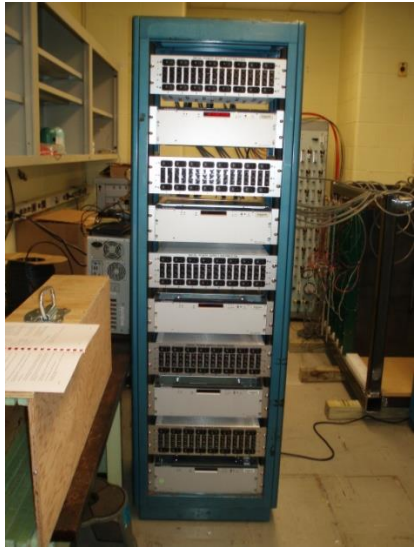

**Figure 16.** Photograph of the low voltage power rack containing five pairs of low voltage power supplies and distribution boxes.

# DATA TAKING WITH THE DHCAL

For the entire period of data taking the threshold to define hits was set to 110 DAC counts, corresponding to about 180 fC, and was not altered. With this threshold setting, DHCAL data were collected in either trigger-less or triggered mode:

a) In trigger-less mode all hits were recorded and time-stamped with a resolution of 100 ns. This mode was particularly useful for measuring the noise rate, collecting cosmic ray events, and for collecting beam data, including events which would not satisfy the trigger conditions, such as events with single photons or neutral hadrons. In addition, this mode was used to monitor the status of the system. Typical monitoring runs lasted 60 s. With an assumed noise rate of 0.5 Hz/cm$^2$, the probability that a given pad records no hits during a 60 s long trigger-less acquisition was negligible. Therefore, any area of the detector showing no hits was a clear indication of a faulty operation, such as due to a loss of high or low voltage or problems with the readout system.

Trigger-less data taking was also used to collect cosmic ray events traversing the whole or any part of the DHCAL. Since the overall noise rate was relatively small and since there was only a negligible rate of accidentals located across more than one layer in a period of four 100 ns time bins, a simple requirement of hits in more than two layers within a 400 ns time-bin selected clean cosmic ray tracks traversing (parts of) the DHCAL.

b) In the test beam, data was collected in triggered data taking mode. In this mode a trigger signal was required to acquire data. In general, for a given trigger signal, hit patterns in seven time bins (each 100 ns long) were recorded. During the test beam campaigns, the timing of the acquisition with respect of the trigger was chosen such that hit patterns in two time bins before the expected hits in the DHCAL were recorded. These provide useful information regarding the noise rate in the detector.

## TESTS WITH COSMIC RAYS

Production RPCs and the final electronic Readout boards were assembled for tests in a cosmic ray test stand. The stand was able to accommodate up to nine chambers, to be read out with 18 Readout boards. Figure 17 shows a photograph of the stand. The RPCs were laid onto strong backs (plywood boards) and were flushed with premixed gas based on the default composition. The data were collected in trigger-less mode and, as a consequence, without hardener. Typical runs lasted five minutes. Events were reconstructed by requiring hits in at least three different layers with timestamps within four time-bins ($\leq$ 400 ns).

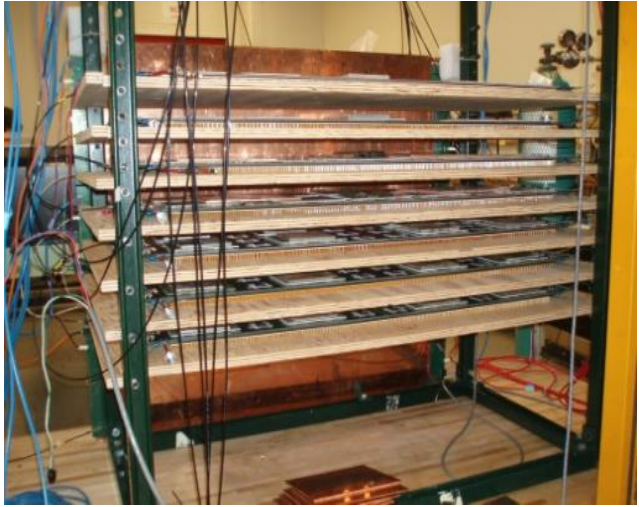

**Figure 17.** Photograph of the cosmic ray test stand.

Similarly to previous analyses of muon data collected with the small-scale prototype [9], the analysis looped over all layers in a given event. The efficiency and pad multiplicity of a given layer was determined from tracks reconstructed using hits in the other layers, in the following named tracking layers. To reconstruct tracks, the hits in each layer were clustered independently from the hits in the adjacent layers. A cluster consisted of hits, which share a common side (closest neighbor algorithm). The center of a cluster was determined as the unweighted arithmetic mean of the positions of the hits belonging to that cluster.

A set of simple cuts resulted in a pure sample of high quality tracks:

a) at least three tracking layers contained hits;
b) tracking layers contained at most one cluster;
c) a straight line was fit in both x/z and y/z (where z is the vertical direction) using the centers of the clusters in the tracking layers. Using a uniform error on the reconstructed position of 1 cm in both x and y, the $\chi^2$ was calculated for both fits. The summed $\chi^2$ values for both fits were required to be less than one;

d) the extrapolated position of the track onto the layer to be investigated was at least 0.5 cm from any edge of its Readout boards.

Figure 18 shows the reconstructed angle of incidence of cosmic rays. The distribution is a convolution of the angular distribution of cosmic rays and the geometrical acceptance of reconstructed tracks.

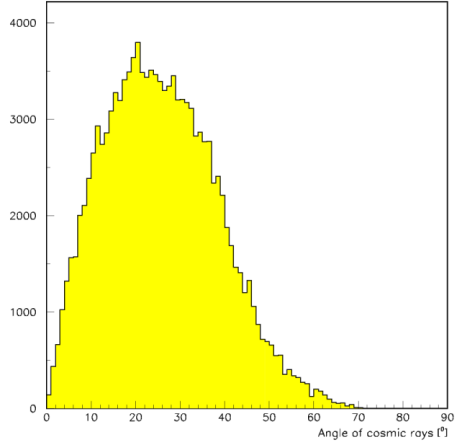

**Figure 18.** Angle of reconstructed cosmic ray tracks. Zero degree corresponds to vertical incidence onto the cosmic ray test stand.

The efficiency, average pad multiplicity and calibration factors as a function of incidence angle are shown in Fig. 19. When operated as a calorimeter, the calibration factors are used to equalize the response from chamber to chamber. Their values are the product of the efficiency and average pad multiplicity. As expected, due to the increased path length of cosmic rays in the gas gap, the values for all three quantities increase with larger angles. The measurements were fit to $2^{nd}$ order polynomials which are seen to describe the angular dependence adequately.

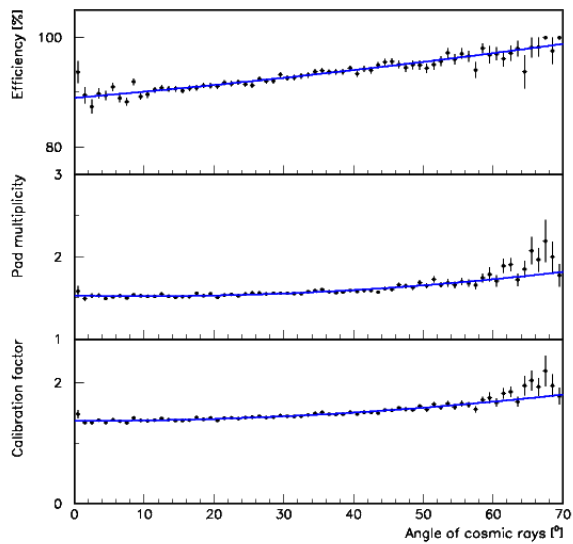

**Figure 19.** Detection efficiency, average pad multiplicity and calibration factors as a function of the angle of incidence of cosmic rays. The blue lines are the results of empirical fits to 2$^{nd}$ order polynomials.

# TESTING IN TEST BEAMS

The DHCAL layers were transported to the Fermilab test beam and were tested in several different configurations. The Fermilab Test Beam Facility FTBF [4] is a world-class facility offering 120 GeV proton beams and secondary beams. The latter can be momentum selected in the range from 1 – 66 GeV/c. At low momenta the secondary beam is dominated by electrons/positrons, at a momentum of about 6 GeV/c equal fractions of electrons/positrons and pions are provided, and at 32 GeV/c and above the fraction of electrons/positrons is at most a few percent.

The beam is delivered once a minute in spills lasting 3.5 seconds. The intensity of the beam can be adjusted to accommodate the needs of different detectors. Due to the limited rate capability of (glass) RPCs [11], the DHCAL requested rates between 50 to 200 Hz, depending on the momentum selection.

The 32 GeV/c secondary beam together with a 3 m long iron beam blocker inserted upstream of the experimental area provided a broadband muon beam. At the location of the DHCAL the muon beam is spread over a large area, of the order of $1 \times 1$ m$^2$. The muon beam was used to measure the response of the RPCs, i.e. the efficiency and average pad multiplicity. The data acquisition was triggered by the coincidence of two $1 \times 1$ m$^2$ scintillator paddles located one in front of the DHCAL and one in the back of the TCMT.

In the period from October 2010 to December 2012, the DHCAL enjoyed five data taking campaigns at FTBF. Table III summarizes the run periods. The secondary beam data include runs at momentum settings in the range of 1 – 60 GeV/c, as well as data taken with the primary 120 GeV proton beam. The April 2011 run included the CALICE Silicon-Tungsten electromagnetic calorimeter [16] placed in front of the DHCAL.

**Table III.** Summary of the data taking campaigns at the Fermilab test beam.

| Run period | Configuration | Combined detector layers | Collected µ events | Collected secondary beam events |
|---|---|---|---|---|
| October 2010 | DHCAL | 38 | 1.4 M | 1.7 M |
| January 2011 | DHCAL+TCMT | 38 + 13 = 51 | 1.6 M | 3.6 M |
| April 2011 | Si-Tungsten ECAL+ DHCAL+ TCMT | 30 + 38 + 14 = 82 | 2.5 M | 5.1 M |
| June 2011 | DHCAL+TCMT | 38 + 14 = 52 | 3.3 M | 2.7 M |
| November 2011 | Minimal absorber | 50 | 0.6 M | 1.3 M |
| **Total** | | | **9.4 M** | **14.4 M** |

The DHCAL layers and their readout electronics were shipped to CERN in 2012, to be tested in conjunction with a tungsten absorber structure. The tests initiated at the Proton-Synchrotron (PS), which provided beams of both polarities between 1 and 10 GeV/c. At the PS particles arrived in spills of 400 milliseconds duration every 45 second. Due to the limited rate capability of RPCs [11], the beam intensity was kept at or below about 500 triggers/spill. Electrons were tagged with a pair of Čerenkov counters, filled with $CO_2$.

After a ten day testing period at the PS, the DHCAL was transferred to the H8 beam line of the Supers-Proton-Synchrotron (SPS), which provided beams in the 10 – 300 GeV energy range. The particles arrived in 9.7 second spills every 45 – 60 seconds. A pair of Čerenkov counters, filled with helium, was used to distinguish electrons from muons/pions up to 25 GeV/c and pions from protons up to 200 GeV/c. Depending on the beam momentum selection, the beam intensity was limited to 250 – 500 triggers/spill. The fraction of pions in the beam could be enhanced by inserting an 18 mm lead foil far upstream, after the momentum selection.

The DHCAL enjoyed four data taking campaigns at CERN. Table IV summarizes the PS and SPS runs.

**Table IV.** Summary of the data taking campaigns at the CERN test beams.

| Test beam | Collected μ events | Collected secondary beam events | Total events |
|---|---|---|---|
| PS | - | 14.5 M | **14.5** |
| SPS | 7.6 M | 18.6 M | **26.2** |
| **Total** | **7.6 M** | **32.1 M** | **40.7** |

To illustrate the imaging capabilities of the DHCAL, Fig. 20 shows a sample of events. Note the absence of isolated hits in the muon event. This is in general the case for muon events and demonstrates the low level of accidental noise hits in the DHCAL. As a consequence, it is understood that the isolated hits in the secondary beam events are part of the electromagnetic/hadronic showers and may be attributed to low-energy neutrals (photons or neutrons) interacting at a distance from the core of the shower.

# CONCLUSIONS

The world's first large scale Digital Hadron Calorimeter (DHCAL) prototype was constructed and assembled in the period from fall 2008 to January 2011. The DHCAL utilizes Resistive Plate Chambers (RPCs) as active elements. The readout is segmented into $1 \times 1$ cm$^2$ pads, read out individually with a 1-bit resolution (=1 threshold = digital readout). Due to its fine segmentation and large size, the prototype holds the world record in channel counts (497,664) both for calorimetry and RPC systems.

The prototype was extensively tested with cosmic rays and in the Fermilab and CERN test beams. It performed according to expectations. A sample of event displays demonstrates the unique imaging and particle identification capabilities of this device. The successful construction and operation of the DHCAL in test beams constitutes an important step in validating this novel concept to hadron calorimetry.

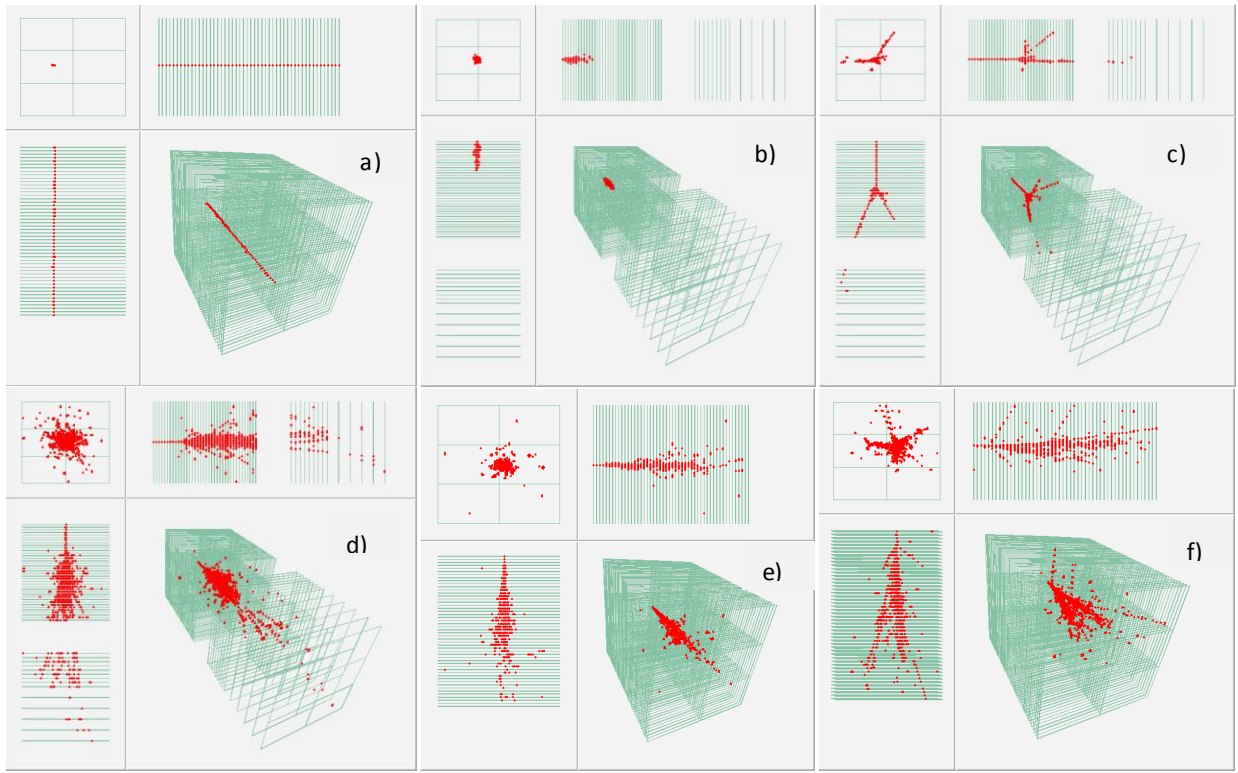

**Figure 20.** Event displays showing different views: clockwise from the left top: x/y, y/z, x/y/z, and x/z. a) Muon track in the minimal absorber stack, b) 8 GeV positron in the DHCAL, c) 8 GeV pion in the DHCAL and TCMT, d) 120 GeV proton in the DHCAL and TCMT, e) 10 GeV positron in the minimal absorber stack, and f) 10 GeV pion in the minimal absorber stack.